\documentclass[12pt,preprint]{aastex}

\usepackage{color}

\shorttitle{{\it Swift} observations of Holmberg~II~X-1}
\shortauthors{Gris\'e et al.}

\begin{document}

\title{X-ray spectral state is not correlated with luminosity in Holmberg~II~X-1}

\author{F. Gris\'e\altaffilmark{1}, P. Kaaret\altaffilmark{1}, H. Feng\altaffilmark{2},
J. J. E. Kajava\altaffilmark{3}, S. A. Farrell\altaffilmark{4}}

\altaffiltext{1}{Department of Physics and Astronomy, University of
Iowa,  Van Allen Hall, Iowa City, IA 52242, USA}
\altaffiltext{2}{Department of Engineering Physics and Center for Astrophysics, Tsinghua University, Beijing 100084, China}
\altaffiltext{4}{Department of Physics and Astronomy, University of Leicester, University Road, Leicester, LE1 7RH, UK}
\altaffiltext{3}{Department of Physics, Astronomy Division, P.O. Box 3000, 90014 University of Oulu, Finland}

\begin{abstract}
The ultraluminous X-ray source (ULX) Holmberg~II~X-1 has been observed over 4 months in 2009/2010 by the {\it Swift} observatory. The source luminosity varied by a factor of up to 14, reaching a maximum 0.3--10 keV luminosity of $\sim 3.0 \times 10^{40}\ \mathrm{erg\ s^{-1}}$.
The spectral properties do not vary much over these 4 months, with only a slight monotonic increase of the hardness ratio with the count rate.
This means that the erratic flaring activity of the source is not associated with spectral changes, as seen in other ULXs. Conversely, comparison with data obtained by {\it Swift} in 2006 shows a completely different picture: while at a luminosity also seen in the 2009/2010 data, the source appears with a hard spectrum.
Thus, it appears that, as in Galactic black hole binaries, spectral states in this ULX are not determined only by the X-ray luminosity.

\end{abstract}


\section{Introduction}

Ultraluminous X-ray sources (ULXs) are point-like X-ray sources located outside the nuclei of galaxies, with luminosities $L_X > 3 \times 10^{39}\ \mathrm{erg\ s^{-1}}$. These apparent luminosities, assuming isotropic emission, are above the Eddington limit of a $20\ M_{\odot}$ black hole. If the Eddington limit applies to these objects then the most likely explanation is accretion onto intermediate-mass black holes (IMBHs) of mass $\sim 10^2 - 10^4\ M_{\odot}$ \citep{Colbert99,Makishima00}.
Alternatively, ULXs may represent a class of super-Eddington emitters and probe a regime in which the accretion rate is much higher than that seen in Galactic black hole binaries (GBHBs; see \citealt{Roberts07} for a recent review).
The reality is probably more complex than a single answer, with some ULXs being possibly slightly more massive than stellar mass black holes ($M_{\mathrm{ULX}} \la 100\ M_{\odot}$) but still experiencing super-Eddington accretion to some degree. In addition, mild beaming may be present \citep{Poutanen07,King09}.

Recently, the {\it Swift} observatory has been used to follow up some of the most luminous ULXs \citep{Kaaret09} and has allowed studies that span a few months to a year with many observations, compared to the previous deeper but sparse {\it Chandra}/{\it XMM-Newton} observations. The results from these observations have shown that their flux varies significantly by factors of 5--10 on timescales of days to weeks \citep{Kaaret09}. In NGC~5408~X-1, a 115 day orbital period has been suggested \citep{Strohmayer09}. Somewhat unexpectedly, most of the ULXs monitored with {\it Swift} to date (Holmberg~IX~X-1, NGC~5408~X-1, and NGC~4395~X-2) do not show pronounced spectral changes \citep{Kaaret09, Vierdayanti10}. The exception is the Hyperluminous X-ray source ($L_X \ge 10^{41}\ \mathrm{erg\ s^{-1}}$; \citealt{Gao03}) HLX-1 \citep{Farrell09} which shows X-ray variability (by a factor of $>20$) but associated with spectral variability \citep{Godet09}. 

Here, we report on a set of new {\it Swift} observations aimed at studying Holmberg~II~X-1 (HoII~X-1 hereafter).
This source is a luminous ULX ($L_{X}$ up to $2\times 10^{40}\ \mathrm{erg\ s^{-1}}$; e.g. \citealt{Feng09}), located in the dwarf galaxy Holmberg~II, at an estimated distance $d=3.39\ \mathrm{Mpc}$ \citep{Karachentsev02}. In optical wavelengths, the ULX is surrounded by the ``foot nebula'' \citep{Pakull02} which is powered by reprocessing of X-rays and has enabled demonstration, using the \ion{He}{2}$\lambda4686$ nebular recombination line, that the true X-ray luminosity is close to that inferred from the X-ray flux assuming isotropic emission, thus excluding strong beaming effects for this source \citep{Pakull02,Kaaret04}. This has been confirmed by an infrared study \citep{Berghea10}, using the [\ion{O}{4}] $25.89\ \mu\mathrm{m}$ emission line.
This ULX has also been studied extensively in the X-ray band \citep{Zezas99,Miyaji01,Dewangan04,Stobbart06,Goad06,Feng09,Gladstone09,Kajava09,Caballero10} and has been seen with 0.3--10 keV luminosities ranging from a few $10^{39}$ to $\sim 2\times 10^{40}\ \mathrm{erg\ s^{-1}}$.

\section{Observations and data analysis}

We obtained 68 observations under {\it Swift} program 90217 (PI: Kaaret) and also
analyzed 6 observations obtained under program 35475. All these observations were carried out using the X-ray Telescope (XRT) instrument in its photon-counting (PC) mode.
The first data set is contiguous and spans 131 days from 2009 December to 2010 April
and the second one comes from observations executed in 2006.
We retrieved level 2 event files from all these observations. Thus, we used
the default data screening parameters as described in the XRT user's guide\footnote{Can be found at \url{http://heasarc.nasa.gov/docs/swift/analysis/}}.
Each observation is composed of one or more snapshots. We analyzed each snapshot
separately, rejecting the ones with an exposure time below 100 s. We were left with
124 snapshots.

We extracted source counts from a circular region with a radius of 20 pixels 
($\sim 47 \arcsec$, corresponding to 90\% of the point-spread function at 1.5 keV) 
and background counts from an annulus with an inner radius of 50 pixels ($\sim 118 \arcsec$)
and an outer radius of 120 pixels ($\sim 283 \arcsec$).
The same regions were chosen to extract the spectra. 
The count rates and spectra were corrected for the loss of flux due to bad pixels
and bad columns. For this, an exposure map was generated for each snapshot and used
to create an auxiliary response file (ARF).
The light curves were binned at 1 day averages to increase statistics. We also calculated a hardness
ratio, defined as the ratio between the net count rate in the 1.5--10 keV band versus the 0.3--1.5 keV band.

We co-added spectra using three different count rate ranges (see Table \ref{tab_fp}) and keeping the 2006 data separate, using the
{\scriptsize  ADDSPEC FTOOL}\footnote{We used FTOOLS 6.9, the latest version being available at \url{http://heasarc.gsfc.nasa.gov/ftools/}}. Thus, we ended up with four co-added spectra, which we refer to throughout the Letter as spectra 1, 2, 3, and 4 (Table \ref{tab_fp}) where spectrum 2 is the spectrum for the 2006 data.
The ARFs were co-added separately using {\scriptsize ADDARF} and weighted
accordingly to their counts.
Response matrix files (RMFs) were taken from the calibration database according to the date of
the observations. Concretely, we used the RMF \verb=swxpc0to12s0_20010101v011.rmf= for the 2006 observations and the RMF \verb=swxpc0to12s6_20070901v011.rmf= for the 2009/2010 observations.
{\scriptsize GRPPHA} was used to bin the spectra with at least 20 counts in each bin.
Finally, we fitted the spectra using {\scriptsize XSPEC} 12.6.0 \citep{Arnaud96}.

\begin{figure*}[tb]
\centerline{\includegraphics[angle=0,width=6.5in]{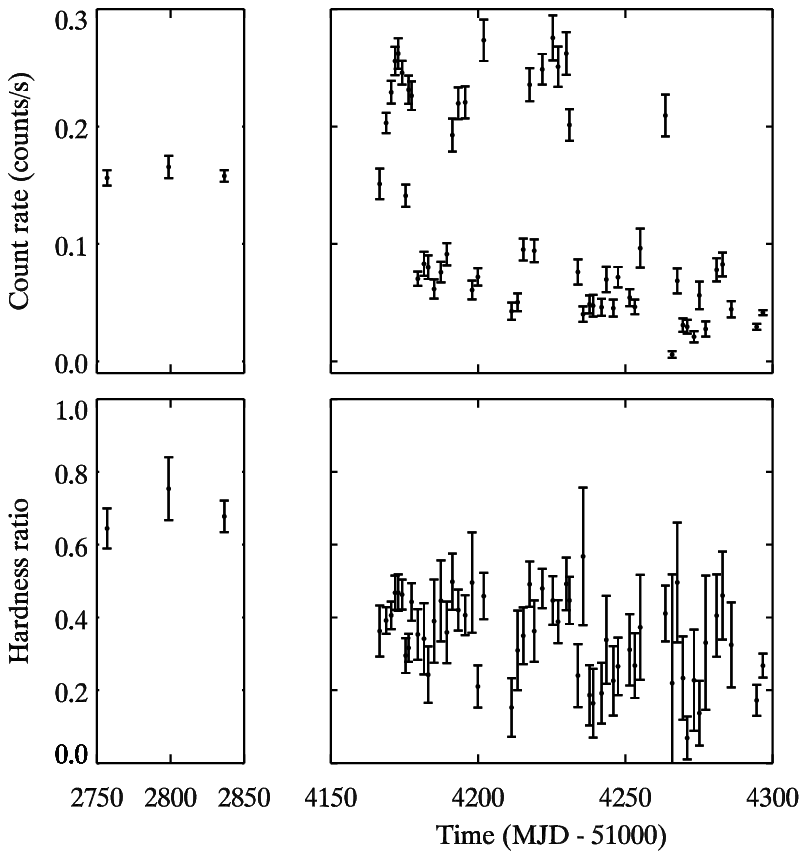}}
\caption{X-ray light curve in the 0.3--10~keV band for Holmberg~II~X-1 (top) and corresponding hardness ratio (bottom).
}
\label{lc}
\end{figure*}


\begin{figure*}[tb] \centerline{\includegraphics[width=6.5in]{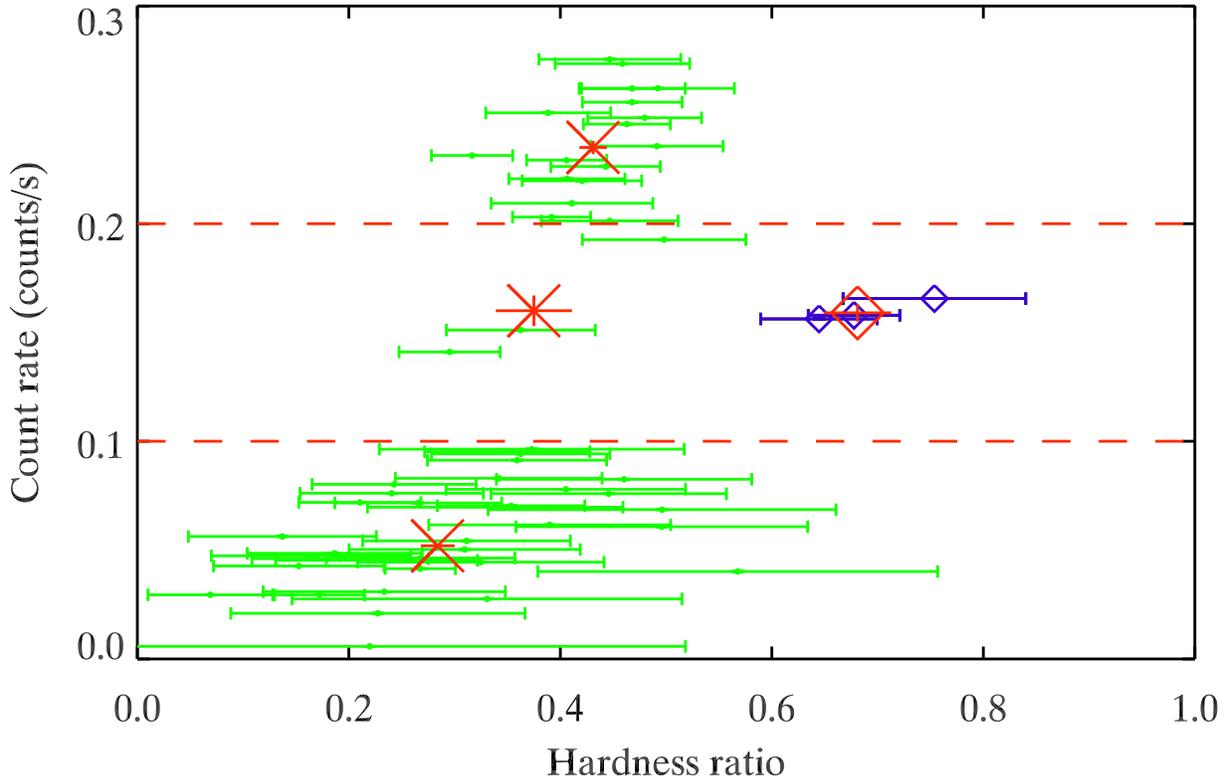}}
\caption{Hardness/intensity diagram for Holmberg II X-1. Each individual observation
from the 2009/2010 observing campaign is plotted as a green bar. The blue diamonds represent the 2006 data.
The red dashed horizontal lines show the count rates ranges used to co-add the spectra. In each count rate range,
we calculated the average hardness ratio, shown by a red cross for the 2009/2010 data or a red diamond for the 2006 data.} 
\label{hi} \end{figure*}

\begin{figure*}[tb] \centerline{\rotatebox{270}{\includegraphics[width=4.5in]{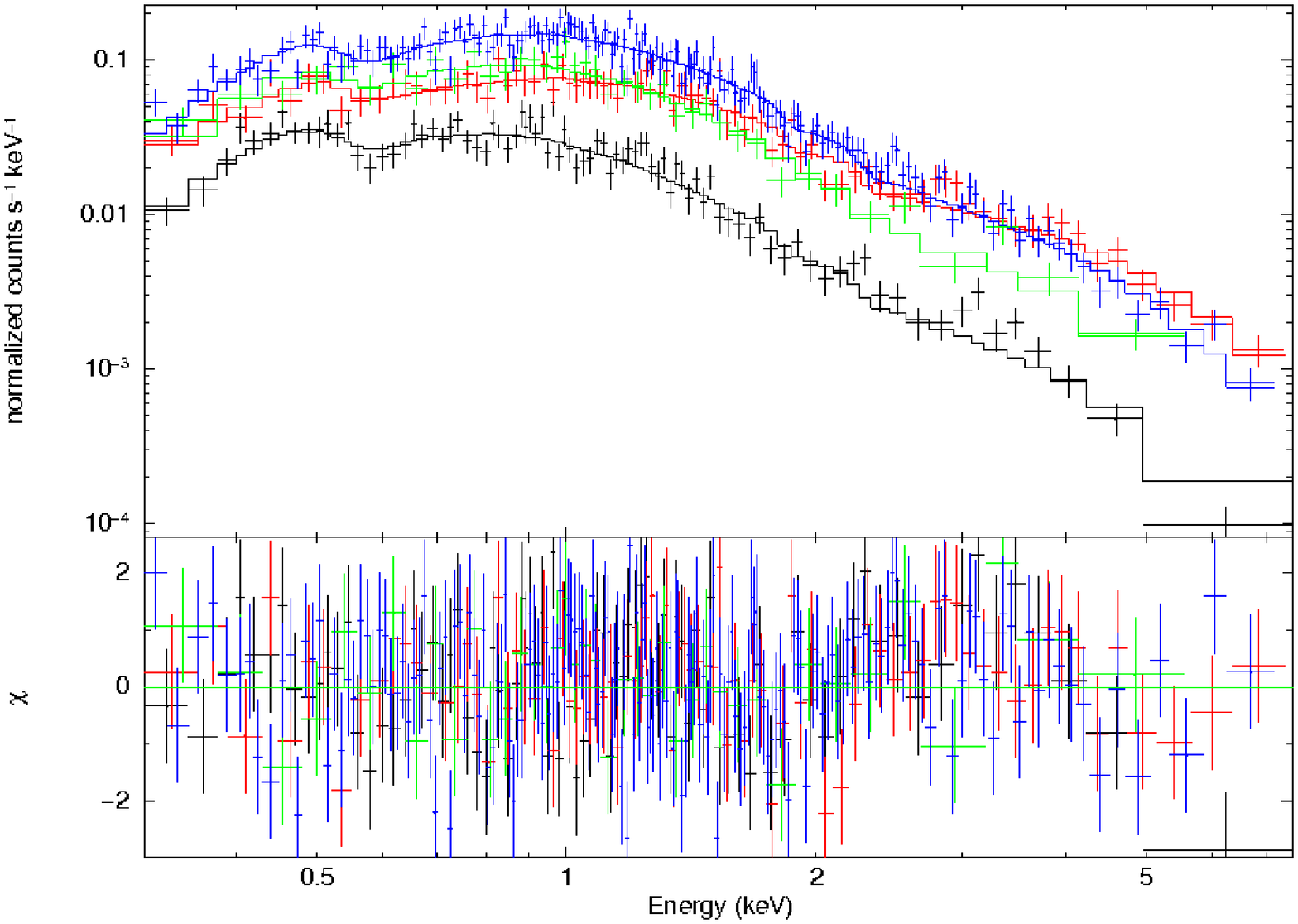}}}
\caption{{\it Swift} XRT best spectral fit using an absorbed multicolor disk plus power-law model.
From bottom to top: spectrum 1 (black), spectrum 2 (red), spectrum 3 (green), and spectrum 4 (blue). The power-law photon indices of spectra 1, 3, and 4 are tied, in accordance with Table \ref{tab_fp}.}
\label{splot} \end{figure*}

\section{Results}

The X-ray light curve of HoII~X-1 (Figure \ref{lc}) shows significant variability,
with count rates ranging from 0.02 to $0.28\ \mathrm{counts\ s^{-1}}$.
We calculated a periodogram from the 0.3--10 keV light curve from the 2009/2010 data set, covering 131 days starting at MJD 55166, using the method
of \citet{Horne86} for periods in the range of 4--65 days. We found no significant peaks in the resulting power spectrum.

The hardness/intensity diagram (HID) is shown in Figure \ref{hi}. The data from the 2009/2010 data set show a slight trend where the hardness ratio increases with the count rate.
This is confirmed by a $\chi^2$ test when we test for the null hypothesis of constant hardness. We get an average hardness of $0.34 \pm 0.01$ and $\chi^2 = 150.6$ for 54 degrees of freedom (DoF), ruling out constant hardness. A linear fit to the 2009/2010 data shows a good correlation between the hardness ratio and the rate, with a slope of $0.98 \pm 0.10$ and a corresponding $\chi^2$ for the fit of 52.2 for 53 DoF.

However, the 2006 observations show a clear offset in hardness with a mean value of $0.68 \pm 0.03$, at a mean count rate of $0.159 \pm 0.004\ \mathrm{counts\ s^{-1}}$. At the same count rate, the mean hardness ratio is $0.38 \pm 0.04$ in the 2009/2010 data.  Clearly, the hardness ratio is not solely a function of the count rate. 

We start the analysis of the spectra (Table \ref{tab_fp} and Figure \ref{splot}) by using the most simple model, i.e., an absorbed power-law continuum.
We fix the galactic \ion{H}{1} column density to $3.4\times 10^{20}\ \mathrm{cm^{-2}}$ \citep {Dickey90} and allow for intrinsic absorption. The $\chi^2$ of the individual fits are statistically acceptable, except for spectrum 1. Spectra 1, 3, and 4 show a consistent extragalactic absorption with a value of $\sim 0.15\times 10^{22}\ \mathrm{cm^{-2}}$ and spectrum 2 shows a lower absorption with $0.05\times 10^{22}\ \mathrm{cm^{-2}}$. The intrinsic column density is thus in reasonable agreement with previous studies \citep[e.g.,][]{Feng09} except the lower value for spectrum 2.
Using the multicolor disk (MCD) blackbody model alone leads to a very bad fit, with a reduced $\chi^2$ above 2.0 for all individual fits.

Adding an MCD component to the power-law model improves the fits significantly in spectra 1 and 4, with $\Delta\chi^2 = 22.5 \;\mathrm{ and }\; 19.4$ for 2 additional DoF. The inner-disk temperature is consistent within the errors between the different spectra, with a value of $\sim 0.20\ \mathrm{keV}$, in agreement with results reported by other authors \citep[e.g.,][]{Feng09}.
After we fitted the power-law index separately, we found that spectra 1 and 4 had the same index within the errors ($\Gamma_1 = 2.80^{+0.23}_{-0.22}\ \mathrm{,}\ \Gamma_4 = 2.57^{+0.14}_{-0.16}$), consistent with the small evolution of the hardness ratio. Spectrum 3 which has fewer counts shows a compatible index but with a larger error ($\Gamma_3 = 2.20^{+0.46}_{-0.69}$). Considering the relatively small variation in hardness ratio over the 2009/2010 data, which may be due to a changing fraction of flux in the soft thermal component, we decided to tie these three indices together. The fit statistics are similar with $\chi^2/{\rm DoF} = 380.2/375$ when the photon indices are independent and $\chi^2/{\rm DoF} = 384.4/377$ when not.  We note that the ``bump'' we see in the residuals of the fit (Figure \ref{splot}) between 2 and 5 keV is likely to be a real feature in the spectra (as seen in a recent {\it XMM-Newton} spectrum; J.J.E Kajava et al., in preparation). This ``bump'' is most apparent in spectrum 1, where the reduced $\chi^2$ of the fit is $\sim 1.2$. More complicated models should be able to model that feature but here we are limited by statistics.

The power-law plus MCD model has been proven to be very effective in the study of black hole X-ray binaries and has been used extensively with the moderate signal-to-noise spectra of ULXs, leading to claims of the presence of IMBHs \citep{Kaaret03,Miller03} due to the presence of a soft excess. But this excess may also arise from the emission of an outflow at a large distance (spherization radius) from the central source \citep{Poutanen07, Kajava09}. Recently, some authors have shown that more physical models could be applied to high-quality spectra of ULXs. \citet{Stobbart06} and \citet{Gladstone09} use a Comptonization model that reflects the coronal emission (and thus replaces the power law) in addition to a disk model. \citet{Caballero10} have applied a model based on a ionized disk reflection and power-law continuum.
We attempted to apply one of the Comptonization models ({\scriptsize DISKPN + COMPTT}, in {\scriptsize XSPEC}) used by \citet{Gladstone09} to our spectra but the fits suffer from huge uncertainties in the fitted parameters. We also summed all the 2009/2010 spectra but were still unable to constrain the model parameters.

It is apparent from the spectra (Figure \ref{splot}) and spectral parameters (Table \ref{tab_fp}) that there are significant changes that are not only a function of the luminosity, as already seen in the HID.
Spectra 1, 3, and 4 share the same power-law index ($\Gamma \sim 2.6$) and inner-disk temperature ($T_{\mathrm{in}} \sim 0.20\ \mathrm{keV}$) within the errors although there is a factor of $\sim 5$ difference in luminosity between spectra 1 and 4. The data from 2006 (spectrum 2) definitely show a harder power-law index ($\Gamma \sim 1.8$), but with the same $T_{\mathrm{in}}$ and a luminosity close to spectrum~3.

\begin{table}[htp]
\caption{Spectral Fit Parameters}
\centering
\tiny
\begin{tabular}{l*{7}{c}r}
\hline
No.\tablenotemark{a}	& Count Rate\tablenotemark{b} & $n_\mathrm{H}$\tablenotemark{c}  & $\Gamma$\tablenotemark{d} & $T_{\mathrm{in}}$\tablenotemark{e} & Flux\tablenotemark{f} & $L_X$\tablenotemark{g} & $f_{X_{\mathrm{MCD}}}$\tablenotemark{h} & $\chi^2$/DoF\tablenotemark{i}\\
     	&	     & ($10^{22}\ \mathrm{cm^{-2}}$) &     &        & ($\times 10^{-12}\ \mathrm{erg\ s^{-1}\ cm^{-2}}$) & ($\times 10^{40}\ \mathrm{erg\ s^{-1}}$) & &\\
\hline
\multicolumn{9}{c}{Power-law} \\
\hline
1	& $< 0.10$          & 0.15$^{+0.02}_{-0.02}$	     & 3.15$^{+0.09}_{-0.09}$		      & ... & 1.3$^{+0.1}_{-0.2}$ & $0.49^{+0.1}_{-0.07}$  & ...  & 122.5 (85) \\
2	& 0.10-0.20 (2006)  & 0.05$^{+0.02}_{-0.02}$	     & 1.98$^{+0.12}_{-0.11}$		      & ... & 5.6$^{+0.4}_{-0.4}$ & $0.97^{+0.05}_{-0.04}$  & ... &  71.8 (82)\\
3	& 0.10-0.20         & 0.15$^{+0.05}_{-0.04}$	     & 2.93$^{+0.25}_{-0.28}$		      & ... & 4.2$^{+0.3}_{-0.6}$ & $1.4^{+0.5}_{-0.3}$	  & ... &  41.6 (38)\\
4	& $> 0.20$          & 0.16$^{+0.02}_{-0.02}$	     & 2.79$^{+0.09}_{-0.09}$		      & ... & 6.7$^{+0.3}_{-0.2}$ & $2.2^{+0.2}_{-0.1}$    & ... &  204.1 (178)\\
\hline
\hline
\multicolumn{9}{c}{Power-law + disk blackbody} \\
\hline
1	& $< 0.10$          & 0.21$^{+0.10}_{-0.07}$ & 2.80$^{+0.23}_{-0.22}$ & 0.15 $^{+0.04}_{-0.03}$ & 1.3$^{+0.1}_{-0.6}$ & $0.67^{+0.7}_{-0.2}$   & $0.51^{+0.49}_{-0.34}$ & 100.0 (83)  \\
2	& 0.10-0.20 (2006)  & 0.05$^{+0.05}_{-0.04}$ & 1.77$^{+0.20}_{-0.26}$ & 0.29 $^{+0.13}_{-0.10}$ & 5.8$^{+0.8}_{-2.7}$ & $1.0^{+0.1}_{-0.1}$    & $0.14^{+0.11}_{-0.10}$ & 66.7 (80) \\
3	& 0.10-0.20         & 0.13$^{+0.11}_{-0.09}$ & 2.20$^{+0.46}_{-0.69}$ & 0.22 $^{+0.10}_{-0.06}$ & 4.5$^{+0.3}_{-4.5}$ & $1.2^{+1.0}_{-0.2}$    & $0.47^{+0.53}_{-0.24}$ & 28.8 (36) \\
4	& $> 0.20$          & 0.18$^{+0.05}_{-0.04}$ & 2.57$^{+0.14}_{-0.16}$ & 0.19 $^{+0.05}_{-0.04}$ & 6.9$^{+0.4}_{-0.8}$ & $2.2^{+0.6}_{-0.4}$    & $0.26^{+0.24}_{-0.12}$ & 184.7 (176) \\
\hline
\hline
\multicolumn{9}{c}{Power-law + disk blackbody with $\Gamma_1 = \Gamma_3 = \Gamma_4$} \\
\hline
1	& $< 0.10$          & 0.17$^{+0.07}_{-0.05}$ & 2.61$^{+0.10}_{-0.12}$ & 0.17 $^{+0.03}_{-0.03}$ & 1.3$^{+0.1}_{-0.6}$ & $0.52^{+0.3}_{-0.1}$   & $0.49^{+0.51}_{-0.26}$  &   \\
2	& 0.10-0.20 (2006)  & 0.06$^{+0.05}_{-0.04}$ & 1.77$^{+0.21}_{-0.27}$ & 0.28 $^{+0.13}_{-0.09}$ & 5.8$^{+1.0}_{-2.9}$ & $1.0^{+0.1}_{-0.1}$    & $0.14^{+0.12}_{-0.10}$ &  \\
3	& 0.10-0.20         & 0.18$^{+0.12}_{-0.06}$ & 2.61 		      & 0.19 $^{+0.06}_{-0.05}$ & 4.3$^{+0.2}_{-1.7}$ & $1.5^{+1.5}_{-0.4}$    & $0.39^{+0.61}_{-0.26}$  &  \\
4	& $> 0.20$          & 0.18$^{+0.05}_{-0.04}$ & 2.61		      & 0.19 $^{+0.04}_{-0.03}$ & 6.8$^{+0.2}_{-1.9}$ & $2.3^{+0.7}_{-0.3}$    & $0.26^{+0.26}_{-0.13}$ &  \\
&&&&&&&&  384.4 (377)
\end{tabular}
\label{tab_fp}
\tablecomments{\tablenotemark{a}Spectrum index used in the text ; \tablenotemark{b}Count rate range used to create the spectrum ; \tablenotemark{c}External absorption column ; \tablenotemark{d}Power-law photon index ; \tablenotemark{e}Inner-disk temperature ; \tablenotemark{f}Absorbed flux (0.3--10 keV) ; \tablenotemark{g}Unabsorbed luminosity (0.3--10 keV) for $D=3.39\ \mathrm{Mpc}$ ; \tablenotemark{h}Fraction of the total unabsorbed flux (0.3--10 keV) in the disk component ; \tablenotemark{i}$\chi^2$ and degrees of freedom. All errors are at the 90\% confidence level. The total number of counts in each of the four combined spectra is respectively 2314, 1940, 943, and 5368 counts for spectra 1, 2, 3, and 4.}
\end{table}

\section{Discussion}

If we attempt a comparison with the spectral states of GBHBs, the power-law index of spectra 1, 3, and 4, which are typical of what has been seen during previous observations ($\Gamma \sim 2.6$) would be consistent with some kind of soft state. The thermal-dominant state as defined by \citet{McClintock06} implies a disk-flux fraction above 75\% and a power-density spectrum with no quasi-periodic oscillations (QPOs) or very weak features. The timing properties would be consistent here since the X-ray variability of HoII~X-1 is weak on short timescales \citep{Goad06}, but the fraction of the flux in the disk component is below 50 \% (with the caveat that the errors on the fractions are huge). Thus, it appears that in 2009/2010 the ULX would be classified as being in the steep power-law state where the power-law component contributes more than 50\% without the presence of QPOs.

Spectrum 2 shows a power-law index ($\Gamma \sim 1.8$) that would be indicative of the hard state. The disk component contributes $\sim 14\%$ of the total X-ray flux which would be consistent with the fraction seen in GBHB in this state. However, the temperature is the same, within errors, in all the spectra. This is unlike the behavior in GBHB, where the disk blackbody usually has a temperature well below that in the soft states \citep{Belloni01,Rodriguez03,Remillard06}.  The properties of the soft thermal component in the spectra of HoII~X-1 are unlike the behavior expected for a disk blackbody. Thus, interpretation of the spectral states of HoII~X-1 in terms of the canonical states at GBHBs in which the thermal emission arises from an accretion disk is not valid.

This particular ULX was only seen once in such a hard-like state \citep[and at a similar luminosity than the {\it Swift} 2006 data;][]{Miyaji01} although it has been observed many times. Looking at the hardness ratios, it appears that HoII~X-1 stayed in this state during the three {\it Swift} observations executed in 2006 although they were separated by $\sim \mathrm{40\ days}$. Since HoII~X-1 has been caught only once before in such a state and the recent coverage of $\sim 130\ \mathrm{days}$ does not show such a hard power-law index, interesting questions are raised such as does this ``transition'' happen rarely and how long does it last.  We note that the evolutionary sequence of \citet{Gladstone09} using Comptonized models is constructed as a function of luminosity and does not include the behavior seen here.  Also, the constancy of the parameters of the disk component, discussed above, would require the corona to vary so as to cancel out the changes in the true disk temperature, which seems unlikely.
It is also worth noting that the few persistent GBHBs behave quite differently compared to the transient ones. Indeed, objects like LMC~X-1, LMC~X-3, or Cyg~X-1 spend most of their time in either the soft state or the hard state with only a few transitions (\citealt{Belloni10}, and references therein). From a strictly spectral hardness point of view, HoII~X-1 also seems to spend most of its time in a soft state. Other ULXs like Holmberg~IX~X-1 seem to be, instead, locked in some kind of hard state \citep{Kaaret09}. We also note that the ULX NGC~1313~X-1 was seen in two different ``states'' at a similar luminosity \citep{Feng06}, hence showing an analogous behavior to HoII~X-1.

On the other hand, the 2009/10 light curve reveals that HoII~X-1 experiences rapid flares that are not tied to spectral changes (i.e., the hardness ratio does not vary much). These flares occur on timescales shorter than the spacing between the observations, $< 2\ \mathrm{days}$ with a mean amplitude of $\sim 0.15\ \mathrm{counts\ s^{-1}}$ (absorbed flux/unabsorbed luminosity amplitudes of $\sim 4.5\times 10^{-12}\ \mathrm{erg\ s^{-1}\ cm^{-2}} \mathrm{/} 1.2\times 10^{40}\ \mathrm{erg\ s^{-1}}$). If we consider all the snapshots without any additional binning, there are two cases that are indicative of a consecutive flux rise and decay in less than half a day each. Other ULXs also show large X-ray variability without significant changes in their spectral hardness \citep{Kaaret09}. The irregular flaring activity that we see here seems very similar to that of Holmberg~IX~X-1 \citep{Kaaret09} and could be also compared to the variability of the GBHB GRS 1915+105, which exhibits rapid flares (on timescales of minutes) that are associated with specific variability patterns \citep{Belloni00} and interpreted as an instability in the radiation pressure dominated part of the disk \citep{Janiuk00}.

We have shown that the ULX HoII~X-1 demonstrates X-ray variability by an average factor of $\sim 5$ on timescales of days to months.  Over 4 months in 2009/2010, its spectral properties did not vary much, with only a slight increase of the hardness ratio associated with an increase in the count rate. During that interval, its spectrum was soft with a steep power-law component and a soft thermal component, similar to that seen during a number of previous observations \citep[e.g.,][]{Goad06,Feng09}. Only the fraction of flux in the soft thermal component has been seen to vary as a function of X-ray luminosity.
This quasi-absence of spectral variability over large changes in flux has been observed in other ULXs \citep{Kaaret09}. But if we add the 2006 data, the HID becomes non-monotonic: the hardness ratio is higher and correspondingly the spectrum shows a harder power law, consistent with the hard state, although its luminosity is close to two of the other spectra. The thermal component appears at a temperature close to that seen in the soft spectra, which argues strongly against interpreting that component as due to disk emission. Thus, while HoII~X-1 appears to exhibit hard versus soft states similar to GBHBs, interpretation of the thermal component in terms of GBHB states is likely invalid.
Even though our spectral analysis is limited by the use of simple and somewhat empirical models, there are spectral changes that are clearly not associated with flux changes. This behavior means that, as with GBHBs \citep{Homan01}, at least two parameters determine the accretion state of this system and that the accretion rate is not the only variable responsible for spectral state changes.

To understand the physics involved, deeper spectra at different count rates and hardness ratios are needed.  The unprecedented coverage possible with {\it Swift} has led to new insights into the behavior of ULXs.  Combined use of {\it Swift} with larger X-ray telescopes could lead to a better understanding of the evolution of the spectral properties of ULXs.

\section*{Acknowledgments}
H.F. is supported by the NSFC under grants 10903004 and 10978001 and by the 973 program 2009CB824800.
J.J.E.K. acknowledges the Finnish Graduate School in Astronomy and Space Physics.
S.A.F. acknowledges STFC funding.
We thank the anonymous referee for their comments which improved this Letter.


\end{document}